\documentclass[aps,prl,twocolumn,showpacs,amsmath,amssymb]{revtex4}
\usepackage{epsfig}
\usepackage{bm}

\begin{document}
\title{Full counting statistics for voltage and dephasing probes}
\author{S. Pilgram$^1$, P. Samuelsson$^2$, H. F\"orster$^3$ and
  M. B\"uttiker$^3$} 
\affiliation{$^1$Theoretische Physik, ETH Z\"urich, CH-8093 Z\"urich,
  Switzerland} 
\affiliation{$^2$Division of Solid State Theory, Lund University, S\"olvegatan 
  14 A, S-223 62 Lund, Sweden}
\affiliation{$^3$D\'epartement de 
Physique Th\'eorique, Universit\'e de Gen\`eve, CH-1211 Gen\`eve 4,
Switzerland}

\date{\today}

\begin{abstract}
We present a stochastic path integral method to calculate the full counting
statistics of conductors with energy conserving dephasing probes and
dissipative voltage probes. The approach is explained for the experimentally
important case of a Mach-Zehnder interferometer, but is easily generalized to
more complicated setups. For all geometries where dephasing may be modeled by
a single one-channel dephasing probe we prove that our method yields the same
full counting statistics as phase averaging of the cumulant generating
function.

\end{abstract}

\maketitle

{\it Introduction} --
Voltage probes are essential elements of many small conductors. In mesoscopic
structures such probes permit to obtain information deep within
the phase-coherent volume \cite{beno,picc,bach}. An ideal {\it voltage probe}
is an infinite impedance terminal with zero net current flow, i.e. any 
electron leaving the conductor through the probe is thermalized by dissipation
and later on fed back into the conductor.  Early theory \cite{mb88,past,ando}
used dissipative voltage probes as simple and successful means to investigate
the transition from quantum coherent conduction to the classical series
addition of resistances. Later on it was realized \cite{Jong96,Langen97} that
theoretically dissipation at a probe can be suppressed by demanding that each
electron exiting into the probe is replaced by an electron incident from the
probe at the same energy.  Such a {\it dephasing probe} can serve as a simple
model to describe dephasing in mesoscopic conduction processes. The power of
this phenomenological approach resides in the fact that the theoretical
modelling may be split into two tasks, first the coherent part of the problem
is treated by the well-developed scattering theory \cite{bee}, second dephasing
is taken into account by evaluating the response of the dephasing probe.

Voltage and dephasing probes also play an important role in the investigation
of noise and current correlations in mesoscopic conductors
\cite{Been92,Jong96,Langen97,bee,Marq1,Forster05,Clerk04}.  As a prominent
example it has been predicted \cite{texi} and recently measured \cite{ober}
that inelastic scattering generated by a voltage probe can render the sign of
cross-correlations positive, while to the contrary anti-bunching of quantum
coherent electrons causes always negative cross-correlations of currents.
Nevertheless, despite the wide use of voltage and dephasing probes
\cite{butrew}, the theory remains in many respects not well developed.

The central aim of this work is to provide a unified description of electrical
transport in the presence of voltage or dephasing probes using the framework
of full counting statistics (FCS) \cite{Naz}.  We employ the stochastic path
integral approach \cite{SPI1} and calculate explicitly the generating
function of the FCS.  Focusing on the dephasing probe, we apply the theory to
a Mach-Zehnder interferometer (MZI).  We compare our results to averaging over
a phase distribution and to classical exclusion models.  For the case of the
MZI we find exact agreement among the models, however this coincidence does
not hold for more complicated scatterers.

\begin{figure}[b]   
\centerline{\psfig{figure=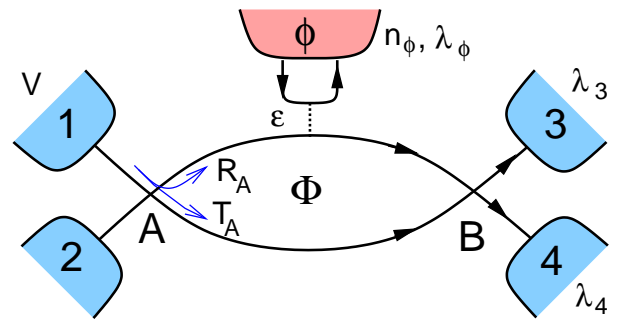,width=8.0cm}}   
\caption{Mach-Zehnder interferometer coupled to a dephasing probe
$\phi$; $\varepsilon$, $\lambda_\phi$, and $n_\phi$ are the coupling strength,
counting field, and the electron distribution of the dephasing probe.}
\label{fig1}
\end{figure}

{\it Dephasing Probe} -- The electronic MZI is shown in
Fig.~\ref{fig1}. It consists of two arms connected to four electronic
reservoirs $1$ to $4$ via reflectionless beamsplitters $A$ and $B$ and to one
dephasing probe $\phi$. Transport in the single mode arms is unidirectional
(indicated by arrows), corresponding to transport along edge-states in the
quantum Hall regime (such a setup was recently realized
experimentally \cite{Ji}).  A bias $eV$ is applied to reservoir $1$, the 
other reservoirs are kept at ground. An Aharonov-Bohm flux $\Phi$ threads the
interferometer.
The transmission (reflection) probability of the beamsplitters is $T_A$ and
$T_B$ ($R_A$ and $R_B$) respectively. The coupling to the probe is
$\varepsilon$, ranging from $0$ for an uncoupled to $1$ for a fully coupled
probe. The scattering matrix $S$ for the total system, including the probe, is
found along the lines of e.g. Ref.~\cite{Chung}. We consider for simplicity
zero temperature, the electron occupation function is thus unity in
reservoir $1$ while zero in reservoirs $2$ to $4$ in the energy interval
$0\leq E\leq eV$ of interest. The occupation function $n_{\phi}(E)$
describes the state of the dephasing probe.

The FCS is the probability distribution $P_c(Q_3,Q_4)=(1/4\pi^2)\int
d\lambda_3d\lambda_4
\exp[-i(\lambda_3Q_3+\lambda_4Q_4)+S_c(\lambda_3,\lambda_4)]$ that
charges $Q_3$ and $Q_4$ are transmitted to reservoirs $3$ and $4$
during the measurement time $\tau$. The FCS is expressed in terms of a
cumulant generating function $S_c(\lambda_3,\lambda_4)$ yielding all
irreducible cumulants by consecutive derivatives $\langle (Q_3)^{n_3}
(Q_4)^{n_4} \rangle = \partial^{n_3+n_4} S_c / (\partial
i\lambda_3)^{n_3} (\partial i\lambda_4)^{n_4}$. To derive $S_c$ we
first obtain the generating function $S_0$ for a fixed value of the
occupation function $n_\phi$ in the dephasing probe. This is achieved
by using the scattering matrix expression introduced by Levitov and
Lesovik \cite{Levitov} $S_0=\frac{1}{h}\int dE \int_0^{\tau}
dt~H_0(\lambda_\phi,n_\phi)$ with
\begin{equation}
H_0=\ln \det \left[1+n\left(\Lambda^{\dagger}S^{\dagger}\Lambda S-1\right)\right]
\end{equation}
and $n=\mbox{diag}(1,0,0,0,n_{\phi})$,
$\Lambda=\mbox{diag}(1,1,e^{i\lambda_3},e^{i\lambda_4},e^{i\lambda_{\phi}})$. 

It is clear that this generating function $S_0$ is not the solution of the
full problem, since current and current fluctuations at the dephasing probe
obtained from derivatives with respect to $\lambda_\phi$ are non-zero. This
violates the defining property of the dephasing probe \cite{mb88}, that for
each energy not only average current but also low-frequency current
fluctuations are zero.  Therefore we introduce fluctuations of the
distribution function $n_{\phi}$ on the slow time scale of $\tau_d$ which is
set by the delay time of particles in the probe. These fluctuations respond in
such a way that the incoming particle current in each energy interval is
compensated exactly by an induced outgoing current.  As a consequence, there
is no charge pile-up in the probe and all carriers end finally up in terminal
$3$ or $4$. This idea can be formalized under the assumption that $\tau_d$ is
much longer than the time spread $\tau_s=h/eV$ of individual electron wave
packets.  This high-voltage assumption allows us to derive the generating
function $S_c$ of the current fluctuations from $S_0$ employing the stochastic
path integral approach \cite{SPI1}.  The condition of charge conservation in
the probe formally enters the total generating function as a factor $i\dot
Q_\phi\lambda_\phi=i\tau_d / h \int dE \dot n_{\phi}\lambda_{\phi}$ where
$Q_\phi$ can be seen as a charge stored in the dephasing probe.  Moreover,
being interested in the charge distribution independent on the particular
realization of $n_{\phi}$ we integrate over the occupation function. As a
consequence the total generating function $S_c$ is expressed as a path
integral
\begin{equation}
\label{path integral}
e^{S_c}=\int {\mathcal D}n_{\phi}{\mathcal D}\lambda_{\phi}\exp
\left\{\int\int_0^{\tau}
\frac{dEdt}{h}  \left[-i\tau_d\lambda_{\phi}\dot n_{\phi}+H_0\right]
\right\}. 
\end{equation}
We evaluate this path integral in the saddle point approximation (Gaussian
corrections are discussed below). The saddle point equations are
\begin{equation}
i\dot n_{\phi}=\frac{1}{\tau_d}\frac{\partial H_0}{\partial \lambda_{\phi}},
\hspace{0.5cm} 
i\dot\lambda_{\phi}=-\frac{1}{\tau_d}\frac{\partial H_0}{\partial n_{\phi}}.
\label{sadpoint}
\end{equation}
In what follows we will only consider the stationary limit, $\tau \gg \tau_d$,
and can therefore omit the time derivatives in Eq.\ (\ref{sadpoint}). The
solutions $\bar \lambda_{\phi}$ and $\bar n_{\phi}$ are readily obtained and
substituted back into $S_c$. This then gives the stationary generating
function ($\psi=2\pi\Phi/(h/e)$ and $N=eV\tau /h$),
\begin{equation}
\bar S_c=N\ln
\left[b\left(1-\frac{\varepsilon}{2}\right)+c\sqrt{1-\varepsilon}\cos\psi+
\frac{\varepsilon}{2}\sqrt{b^2-c^2}\right],
\label{genfcn}
\end{equation}
the main result of this work. We note that the first and second derivatives of
$\bar S_c$ reproduce known results for current and noise calculated with the
dephasing probe \cite{Seelig,Marq1,Chung,Marq2} and that the generating
function becomes flux independent in the limit of strong dephasing,
$\varepsilon\rightarrow 1$.  The first term in $\bar S_c$ is proportional to
$b=(T_AT_B+R_AR_B)e^{i\lambda_3}+(T_AR_B+R_AT_B)e^{i\lambda_4}$ representing
the classical contribution due to particles which go either along the upper or
the lower arm, the second term is proportional to
$c=2\sqrt{R_AT_AR_BT_B}\left(e^{i\lambda_4}-e^{i\lambda_3}\right)$ giving the
coherent quantum interference contribution.  The last term proportional to
$\sqrt{b^2-c^2}$ has an exchange, or two-particle interference origin: the
dephasing probe gives rise to two-particle processes such as $1\rightarrow
3,\phi\rightarrow \phi$ and $1\rightarrow \phi,\phi\rightarrow 3$ which are
indistinguishable.

To determine the range of validity of the saddle point approximation,
we introduce fluctuations of the occupation number $n_\phi =
\bar{n}_\phi + \delta n_\phi(t)$ and of the Lagrange multiplier
$\lambda_\phi = \bar{\lambda}_\phi + \delta \lambda_\phi(t)$ and
expand the exponent of Eq.\ (\ref{path integral}) up to second order
in the fluctuations. Carrying out the gaussian integrals we obtain the
contribution
\begin{eqnarray}
\delta^2 S_c &= &-\frac{\tau_s}{\tau_d}N
\frac{\varepsilon}{4R_A T_A \exp(\bar S_c/N)}\\
	&&\hspace{-6mm}\left|2
  R_A(R_Be^{i\lambda_3}+T_Be^{i\lambda_4})-b+(T_A-R_A) 
  \sqrt{b^2-c^2} \right|\nonumber
\end{eqnarray}
The saddle point approximation is thus valid if $\tau_s / \tau_d \ll 1$. Note
that the delay time can be expressed in terms of a density of states
$\tau_d=hN_F$ of the dephasing probe. One may thus equivalently state that our
approximation is valid if the number of states in the probe participating in
transport is large, $N_FeV \gg 1$. It should also be emphasized that the path
integral calculation is easily extended to more than one probe: it is
sufficient to replace $n_\phi$ and $\lambda_\phi$ by corresponding vector
quantities.

\begin{figure}[b]   
\centerline{\psfig{figure=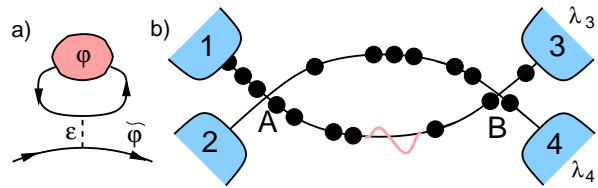,width=7.8cm}}   
\caption{a) Phase averaging: Electrons aquire a total phase factor
$e^{i\tilde \varphi}$ from multiple reflections (amplitude
$e^{i\varphi}$) at a coherent scatterer. b) The exchange (classical
ball) model for equal (black) and different (pink) arm lengths.}
\label{fig2}
\end{figure}

{\it Phase averaging} -- It is interesting to compare the dephasing probe
approach with the most elementary model for dephasing, a phase
averaging. Consider the upper arm of the MZI coupled with strength
$\varepsilon$ not to a dephasing probe, but to an elastic (coherent) scatterer
reflecting with a phase factor $e^{i\varphi}$ (see Fig.\ \ref{fig2}a).  Guided
by random matrix theory~\cite{bee}, we assume that the phase $\varphi$ is
uniformly distributed and therefore simply average the cumulant generating
function over the scattering phase~\cite{footnote phase}.  An electron
traversing the upper path can enter the scatterer and after multiple internal
reflections continue on its path with an additional total phase factor
$e^{i\tilde \varphi}$. The electron phase $\tilde \varphi$ is related to the
scattering phase by $\varphi=\tilde\varphi+\pi+2\arctan[\sqrt{1-\varepsilon}
\sin\tilde\varphi/(1-\sqrt{1-\varepsilon} \cos\tilde\varphi)]$.  In terms of
the new variable $\tilde \varphi$ the phase average becomes
\begin{equation}
\langle S_c \rangle_\varphi
=
N
\int_0^{2\pi} d\tilde{\varphi} f(\tilde{\varphi}) 
 H_0(\psi+\tilde{\varphi})|
_{\varepsilon=0}, 
\end{equation}
where the $2\pi$ periodic distribution of
$\tilde\varphi$ is given by
\begin{equation}
	f(\tilde\varphi)=\frac{1}{2\pi}\frac{d\varphi}{d\tilde\varphi}=
	\frac{1}{2\pi}\frac{\varepsilon}{2-\varepsilon+
	2\sqrt{1-\varepsilon} \cos(\tilde{\varphi})}.
\end{equation}
The outcome of the phase average agrees exactly with the result\
(\ref{genfcn}) obtained from the dephasing probe model.  Therefore one can
argue that the dephasing probe acts like a disordered mirror that reflects
with the phase distribution $f(\tilde\varphi)$. We remark that a microscopic
model for the scatterer is a single mode chaotic cavity with a dwell time much
longer than $\tau_s$ (recently, the influence of dephasing on noise was
modelled using a many mode chaotic cavity \cite{Been05b}). It is possible to
show that the agreement between the dephasing probe and phase averaging holds
for an arbitrary mesoscopic conductor connected to a single probe. For a conductor characterized by a scattering matrix $S$ of size $M+1$ (last index denotes the probe terminal) the generating function is given by
\begin{equation}
S_c=N\ln\left[{\mathcal A}+\sqrt{{\mathcal A}^2-4\mathcal B}\right]
\end{equation}
with ${\mathcal A}=F_{00}+F_{11}-F_{10}$ and ${\mathcal
B}=F_{10}F_{01}-F_{00}F_{11}$ where $F_{\alpha\beta}=\det[1+\bar
n_{\alpha}(\Lambda_{1}^{\dagger}S^{\dagger}\Lambda_{\beta}S-1)]$ with $\bar
n_{\alpha}=\text{diag}(n_1,n_2,...n_M,\alpha)$ and
$\Lambda_{\alpha}=\text{diag}(e^{i\lambda_1},e^{i\lambda_2},
...e^{i\lambda_M},\alpha)$.  Note that for any geometry containing two or more
probes, the dephasing probes destroy interference between paths that pass a
given set of probes in different chronological order. Such an interference is
not removed by phase averaging.  Therefore, it is generally not possible to
describe the effect of many dephasing probes with phase averaging.

{\it Classical exclusion models} -- The FCS in the limit of complete dephasing
($\varepsilon=1$) may also be understood from a completely classical point of
view.  In exclusion (classical ball) models electrons enter the
interferometer through terminal $1$ as a regular noiseless stream of classical
particles (see Fig.\ \ref{fig2}b). Time and space are discretized and in each
time step the particles in the interferometer arms propagate one site
forward. At the beamsplitters $A$ and $B$ the particles scatter with the same
probabilities, $T_A,R_A,T_B$ and $R_B$, as in the quantum case. The Pauli
principle is introduced by hand by excluding scattering events which would
lead to two electrons occupying the same site.

To calculate the noise of the MZI in the classical limit ($\varepsilon=1$),
Marquardt and Bruder \cite{Marq1} consider an exclusion model with both arms
of the interferometer having equal length (the same number of
sites). Consequently, two electrons can not arrive simultaneously at
beamsplitter $B$ and the Pauli principle never comes into play. The generating
function for this model is given by
\begin{equation}\label{equalarms}
	S_c=N\ln b.
\end{equation}
Surprisingly, there is a disagreement between the exclusion model result and
the dephasing probe (and phase averaging) result in Eq.\ (\ref{genfcn}). The
fact that the dephasing probe model does not reproduce a seemingly obvious
classical result lead the authors of Refs.\ \cite{Marq1} to strongly question
the reliability of the dephasing probe.

As it turns out, this contradiction can be resolved by considering a model
with unequal arm lengths \cite{Marq1}. Such a model takes into account that the dephasing
probe effectively delays the particle a time $\sim \tau_d$. Particles that
enter the interferometer at different times might thus arrive at the second
beamsplitter B simultaneously and consequently, the Pauli principle comes into
play.  The generating function for an interferometer with finite arm length
difference can be calculated with known methods, see e.g.\ \cite{roche} and is
found to be
\begin{equation}
	S_c=N\ln[b/2+1/2\sqrt{b^2-c^2}]
\end{equation}
which completely agrees with Eq.\ (\ref{genfcn}) in the limit of
$\varepsilon=1$. The 
term $\sqrt{b^2-c^2}$ identified above as an exchange term 
is in the
classical ball model a direct consequence of the Pauli principle. The
dephasing probe model thus captures correctly two essential
ingredients of the FCS in the presence of dephasing; it is able to
destroy phase information in the fermionic wavefunction and to keep at
the same time its antisymmetry, the Pauli principle.

{\it Dephasing versus voltage probes} -- While in a dephasing probe the
occupation number $n_\phi$ fluctuates to conserve current at each energy, in a
voltage probe $n_\phi$ is Fermi distributed and the voltage $V_\phi$
fluctuates to conserve the total, energy integrated current
\cite{Been92,texi,ober,Kindermann04}. The generating function in the limit of
zero temperature is
\begin{equation}
	S_V=\frac{\tau}{h}\left[\int_0^{eV_\phi}\!dEH_0|_{n_\phi=1}+\int_{eV_\phi}^{eV}\!dEH_0|_{n_\phi=0}\right].
\end{equation}
In analogy to the procedure for the dephasing probe, we get saddle point
equations for the long time limit $dS_V/d\lambda_{\phi}=0$ and
$dS_V/dV_{\phi}=0$.  Importantly, for a single probe in the
linear voltage regime the saddle points for dephasing and voltage
probes are equivalent, i.e. the FCS does not depend on the presence of
dissipation.

Differences between dephasing and voltage probes become visible when two or
more probes are attached.  As an illustrative example, we consider the MZI
with a second probe attached to the lower arm of the interferometer, similar
to Fig.\ \ref{fig1}. The corresponding unconstraint generating function is
nonlinear in occupation numbers and counting variables of the probes, leading
to more complicated saddle point equations. We limit our discussion to the
first two cumulants, mean current and noise in terminal $3$, which already
differ for the two probe types.

We first consider the case where the two probes are {\it
disconnected}, characterized by independent distribution function
$n_{\phi 1}$ and $n_{\phi 2}$ or voltages $V_{\phi 1}$ and $V_{\phi
2}$ respectively. For the dephasing probe the current and noise are
obtained from the single probe result, Eq.\ (\ref{genfcn}) by
replacing $\sqrt{1-\varepsilon}$ with $1-\varepsilon$.  The same
result is found for the voltage probe current but the noise differs by
a term proportional to $\varepsilon^2$. Interestingly, the dephasing
probe result is also obtained from phase averaging by considering, in
analogy to the single probe case, two uniformly and independently
distributed phases $\varphi_1$ and $\varphi_2$, one for each
probe. This equivalence, a result of the simple MZI geometry, does not
hold for more complicated systems.

For the case with the two {\it connected probes}, characterized by a single
distribution function $n_{\phi}$ and a single voltage $V_{\phi}$
respectively, the situation is quite different. A particle entering
the probe in e.g.\ the upper arm can be reemitted in the lower arm,
thus only the sum of the currents into the two probes is
conserved.
In the limit of complete dephasing, $\varepsilon=1$, the problem simplifies
considerably: all particles injected
from $1$ are first emitted into one 
of the two probes and then reemitted again before exiting into
terminal $3$ or $4$. The mean current is simply $I=(e^2V/h)/2$,
independent of the beamsplitter scattering probabilities. For a
voltage probe the noise is suppressed to zero because the
outgoing streams from the two probes are noiseless, due to unit
occupation in the probes up to energy $eV/2$. For a dephasing probe
the distribution function is instead a two-step function with
$n_\phi=1/2$ in the  energy interval $[0,eV]$. Thus, the streams
incident on $B$ are occupied with probability $1/2$, leading to a noise
$S_{deph}=(2e^3V/h)/8$, again independent on beamsplitter scattering
probabilities.  Note that there is no natural extension of the phase
average to the case with two connected probes.

{\it Conclusion} -- We have extended the discussion of voltage probes
and dephasing probes for current and noise to the level of full
counting statistics, valid for the case that the dephasing mechanism
is slow and averages over many electron wave packets.  For conductors
connected to a single dephasing probe we find that there exists a
phase distribution such that the phase average of the generating function
of FCS is identical to the generating function of the conductor with
the dephasing probe. Contrary to statements in the literature, this
proves that dephasing probes correctly account for the Pauli principle
and exchange effects. The stochastic path integral approach, which
correctly describes exchange effects in the multichannel limit
\cite{SPI1}, can readily be extended to the multiple (or
multichannel) probe case. No such simple extension is possible for
phase averaging.  The physics of voltage probes, an essential element
of electrical transport, and the closely related dephasing probe
approach will likely also in the future remain an important element in
discussions of the statistics of the conduction process.

We thank E. V. Sukhorukov for discussions. This work is supported by MaNEP 
and the Swiss and Swedish NSF.


\vspace{-.5cm} 
\begin{thebibliography}{02}

\vspace{-.5cm} 
\bibitem{beno}    
A. D. Benoit, S. Washburn, C. P. Umbach, R. B. Laibowitz, and R. A. Webb,
Phys. Rev. Lett. {\bf 57}, 1765 (1986). 
\bibitem{picc} 
R. de Picciotto, H. L. Stormer, L. N. Pfeiffer, K. W. Baldwin, K. W. West, 
Nature {\bf 411}, 51 (2001).             
\bibitem{bach} 
B. Gao, Y. F. Chen, M. S. Fuhrer, D. C. Glattli, A. Bachtold,
Phys. Rev. Lett. {\bf 95}, 196802 (2005). 
\bibitem{mb88}    
M. B\"{u}ttiker, 
IBM J. Res. Develop., {\bf 32}, 63-75 (1988). 
\bibitem{past}     
J. D'Amato and H. M. Pastawski, Phys. Rev. B {\bf 41}, 7411 (1990).           
\bibitem{ando}     
T. Ando, Surface Science {\bf 361/362}, 267 (1996).
\bibitem{Jong96} 
M. J. M. de Jong, and C. W. J. Beenakker, Physica A {\bf 230}, 219 (1996).
\bibitem{Langen97} 
S. A. van Langen, and M. B\"{u}ttiker, Phys. Rev. B \textbf{56}, R1680 (1997).
\bibitem{bee}    
C. W. J. Beenakker, Rev. Mod. Phys. {\bf 69}, 000731 (1997). 
\bibitem{Been92} 
C. W. J. Beenakker, and M. B\"uttiker, Phys. Rev. B {\bf 46}, 1889 (1992).   
\bibitem{Marq1} F. Marquardt and C. Bruder, Phys. Rev. Lett. 
\textbf{92}, 56805 (2004); Phys. Rev. B \textbf{70},
125305 (2004). 
\bibitem{Clerk04}
A. A. Clerk and A. D. Stone, Phys. Rev. B {\bf 69}, 245303 (2004);
E. Prada, F. Taddei, and R. Fazio, Phys. Rev. B {\bf 72}, 125333 (2005).
\bibitem{Forster05}
H. F\"orster, S. Pilgram and M. B\"uttiker, Phys. Rev. B \textbf{72}, 075301
(2005).
\bibitem{texi}    
C. Texier and M. B\"uttiker, Phys. Rev. B {\bf 62}, 7454 (2000). 
\bibitem{ober}
S. Oberholzer, E. Bieri, C. Schoenenberger, M. Giovannini, J. Faist, 
Phys. Rev. Lett. {\bf 96}, 046804 (2006). 
\bibitem{butrew} 
Ya. M. Blanter and M. B\"{u}ttiker, Phys. Rep. {\bf 336} 1 (2000).
\bibitem{Naz} For a recent review see {\it Quantum noise in Mesoscopic
  Physics}, edited by Yu. V. Nazarov (Kluwer, 2003). 
\bibitem{SPI1}
S. Pilgram, A. Jordan, E. V. Sukhorukov and M. B\"uttiker,
Phys. Rev. Lett. {\bf 90}, 206801 (2003);
A. N. Jordan, E. V. Sukhorukov and S. Pilgram, J. Math. Phys. {\bf 45}, 4386
(2004); for related work see also 
D.~B. Gutman, A.~D. Mirlin, and Y. Gefen,
Phys. Rev. B {\bf 71},85118 (2005).
\bibitem{Ji} 
Y. Ji, Y. Chung, D. Sprinzak, M. Heilblum, D. Mahalu and H. Shtrikman, Nature
\textbf{422}, 415 (2003).
\bibitem{Chung}
V. S.-W. Chung, P. Samuelsson, M. B\"uttiker,
Phys. Rev. B {\bf 72}, 125320 (2005).
\bibitem{Seelig} 
G. Seelig, M. B\"{u}ttiker, Phys. Rev. B {\bf 64}, 245313 (2001).
\bibitem{Marq2}
F. Marquardt, Europhys. Lett.  {\bf 72}, 788 (2005).
\bibitem{Levitov} 
L. S. Levitov and G. Lesovik, JETP Lett. {\bf 58}, 230 (1993).
\bibitem{footnote phase} The phase average $\langle S_c \rangle_\varphi$
excludes modulation contributions which appear in an experiment with external
phase fluctuations. There one rather measures the logarithm of the averaged
moment generating function $\ln \langle e^{S_c} \rangle_\varphi$ which is
discussed in Ref.\ \cite{Forster05}. The two averages coincide only to linear
order in the voltage $eV$.
\bibitem{Been05b}
C. W. J. Beenakker and B. Michaelis, J. Phys. A: Math. Gen. {\bf 38}, 10639 (2005).
\bibitem{roche} P.-E. Roche, B. Derrida and B. Doucot, Eur. Phys. J. B 
\textbf{43}, 529 (2005).
\bibitem{Kindermann04} FCS in the presence of voltage fluctuations in networks
of two-terminal conductors is discussed in M.  Kindermann, Yu. V. Nazarov, and
C. W. J. Beenakker, Phys. Rev. B {\bf 69}, 035336 (2004).
\end{thebibliography}
\end{document}